\documentclass[10.5pt,twocolumn]{article}
\usepackage[left=13mm, top=25mm, bottom=25mm, right=13mm]{geometry}
\usepackage{xcolor}
\usepackage{sectsty}
\usepackage{graphicx}
\usepackage{amsmath}
\usepackage[nomove,sort]{cite}
\usepackage{comment}
\usepackage{textcomp}
\usepackage{lineno}
\makeatletter
\renewcommand{\@biblabel}[1]{#1.}
\makeatother

\sectionfont{\color{black}}
\subsectionfont{\color{blue}}

\begin{document}
\title{Macro-coherent radiative emission of neutrino pair
between parity-even atomic states}

\author{
M.~Tashiro{$^1$}\thanks{e-mail: tashiro046@toyo.jp},
B.~P.~Das{$^2$},
J.~Ekman{$^3$},
P.~J\"onsson{$^3$},
N.~Sasao{$^4$} and 
M.~Yoshimura{$^4$} \\[5mm]
$^1$Department of Applied Chemistry, Toyo University, Kujirai 2100, \\
           Kawagoe, Saitama 350-8585, Japan \\ 
$^2$Department of Physics,
           Tokyo Institute of Technology, 2-12-1-H86 Ookayama, \\ Meguro-ku, Tokyo 152-8550, Japan \\ 
$^3$Department of Materials Science and Applied Mathematics,\\ Malm\"o University, 20506 Malm\"o, Sweden \\ 
$^4$Research Institute for Interdisciplinary Science,
           Okayama University, 700-8530 
}

\date{\today}
\maketitle

\abstract{
A new scheme to determine the neutrino mass matrix is proposed using
atomic de-excitation between two states of a few eV energy spacing.
The determination of the smallest neutrino mass of the order of 1 meV 
and neutrino mass type, Majorana or Dirac,
becomes possible, if one can coherently excite more than 1 gram of 
atoms using two lasers.
}\\[5mm]
\noindent
keywords:{neutrino mass matrix, CP violation, Majorana fermion}

\vspace{0.5cm}

\section{Introduction}
\label{intro}

Despite of remarkable success in oscillation experiments \cite{pdg}, neutrino physics
has left more conundrums than what it has discovered and already established:
major conundrums are
how small the lightest neutrino mass is
and whether neutrinos are Majorana or Dirac type of fermions (the Majorana/Dirac distinction).
These two items are linked to  important issues in macro- and micro-
worlds:   how our matter dominated universe was created (the problem of
baryon asymmetry of the present universe)
and the unified theory beyond the standard $SU(3) \times SU(2)\times U(1)$ gauge theory. 
In this respect it is crucially important to invent new experimental schemes of neutrino physics
beyond  the established technology based on nuclear targets, since they release too large energy of a few to several MeV, 
much larger than expected small neutrino masses in the sub-eV range.

One possible scheme is to use a target system of available energy closer to the
expected range of neutrino masses (0.1 $\sim 10^{-3}$) eV.
It was recently suggested by a few of the authors 
that isolated atomic systems might be
a solution provided that an effective mechanism of coherent amplification
of event rates is realized \cite{renp overview}.
Translated to the single atomic rate,
this scheme enhances the rate by the total number of phase-coherent atoms. 
One of the problems in the proposed scheme is however scarce candidate atoms:
the best candidate in terms of rate so far found is Xe gas, but due to a large required laser energy
(in sum $\sim$ 8 eV)
the Xe scheme may encounter  a great challenge of copious ionization loss with present
laser technology.

A related important question is how many target atoms are prepared:
high-density atomic target in solids is an obvious choice, but various relaxation processes
arising in solids may destroy coherence.
A possible solution is to encapsulate atoms, 
which may maintain  isolated features of atoms in vacuum, yet may not possess
relaxation phenomena present in ordinary solids.
We propose here a new scheme to evade this difficulty.

We use the natural unit of $\hbar = c = 1$ throughout the present paper unless otherwise stated.

\section{Radiative emission of neutrino pair: use of new interaction pieces}
\label{sec:1}

We consider electroweak process of atomic de-excitation
from a metastable excited state $| e \rangle$ to the ground state $|g \rangle$, 
$| e \rangle \rightarrow |g \rangle + \gamma + \nu \bar{\nu}$, with
$\gamma$ a photon and $ \nu \bar{\nu} $ a neutrino-pair.
This process is described by
the second order perturbation  theory of combined electric dipole (E1)  photon emission 
and four-Fermi type of weak process of neutrino pair emission. 
The coherently amplified process of this type
has been termed RENP (Radiative Emission of Neutrino Pair) \cite{renp overview}.
The phase-coherence over a macroscopic body of atoms
has been experimentally verified in second order QED process to achieve
a rate enhancement of the order of $10^{18}$
in agreement with theoretical expectation \cite{psr exp}.
With macro-coherence of phases, the probability amplitudes obey
\begin{eqnarray}
&&
\sum_a e^{i (\vec{p}_{eg} - \vec{k}_{\gamma} - \vec{p}_1 - \vec{p}_2)\cdot \vec{x}_a} {\cal A}(\vec{x}_a )\simeq
\nonumber \\
&& \hspace{10mm}
 \frac{N}{V} (2\pi)^3 \delta^3 (\vec{p}_{eg} - \vec{k}_{\gamma} - \vec{p}_1 - \vec{p}_2 )  {\cal A}_0
\,,
\end{eqnarray}
where $\vec{p}_{eg} \,, \vec{k}_{\gamma} $ are  wave vectors imprinted at laser excitation
and emitted photon, and $\vec{p}_{i}$ (i = 1,2) are momenta of the emitted neutrino-pair.
Emergence of a macroscopic quantity, the atomic number density $n= N/V$,
is ensured provided phases of atomic wave functions ${\cal A}(\vec{x}_a ) =  {\cal A}_0$ are 
common and not random over sites.

Since we use in the present work new pieces of electron-neutrino interaction,
we shall recapitulate results of the established  electroweak theory of three flavors
extended to accommodate finite neutrino masses.
Adding neutral current 
 and  charged current interactions after Fierz transformation,
the neutrino pair emission is described by the four-Fermi type interaction hamiltonian density,
\begin{eqnarray}
&& 
\frac{G_F}{\sqrt{2}}\, \Sigma_{ij}\,\Sigma_{\alpha}\,
\bar{\nu}_i \gamma_{\alpha}  (1 - \gamma_5)\nu_j \,
\bar{e}\left( \gamma^{\alpha} c_{ij} - \gamma^{\alpha} \gamma_5 b_{ij}
\right) e
\,,
\label {weak cc}
\\ && \hspace{10mm}
c_{ij} = U_{ei}^* U_{ej} - \frac{1}{2} ( 1- 4 \sin^2 \theta_w) \delta_{ij}
\,, 
\\ && \hspace{10mm}
b_{ij} = U_{ei}^* U_{ej} - \frac{1}{2}\delta_{ij}
\,,
\end{eqnarray}
$\nu_i (i=1,2,3)$ denoting neutrino mass
eigenstates of masses $m_i$.
The $3\times 3$ matrix element $U_{ai}$ ($a=e, \mu, \tau$) describes
neutrino mass mixing.
Experimentally, $1 - 4\sin^2 \theta_W \sim 0.046 \pm 0.00 64$.

There may be two RENP possibilities depending on which electron operators are used:
(1) spatial component of axial vector current in Eq.(\ref {weak cc}), (2) time component of axial vector
or spatial component of vector current.
The temporal part of vector current does not contribute much due to orthogonality of wave functions,
which is justified when electrons move in atoms non-relativistically.
The dominant atomic electron contribution arises from the spatial part of axial vector
given by the spin operator, $e^{\dagger} \vec{\alpha}\gamma_5 e = e^{\dagger} \vec{\Sigma} e$
($\vec{\Sigma}  $ being the $4\times 4$ Pauli matrix).
This is dominant since there are no mixing between large and small components
of Dirac 4-spinor wave functions.
Combined with electric dipole (E1) photon emission, the angular momentum and parity change
of atomic electrons is of the type, $\Delta J^P = 0^-, 1^-, 2^- $, and hence
may be called parity-odd RENP.
The best initial atomic state in terms of rates  is Xe ($J^P = 2^-)$.
There is however a problem due to a large excitation energy $\sim 8.3\,$eV,
such as a close elusive ionization level.

On the other hand, there are many atomic candidates for parity-even transitions
between states 
having excitation energies much less than in the parity-odd transition case.
For this purpose we shall study parity-even RENP which have smaller rates, but
may have better chances to be realized in simpler experiments. 
The sub-dominant contributions arise from either of two ways
of taking electron operators: the spatial part of vector, 
$\bar{e} \vec{\gamma} e = e^{\dagger} \vec{\alpha} e$, and zero-th component of axial vector, 
$e^{\dagger}\gamma_5 e$.
Both of these are suppressed by the velocity of atomic electrons, 
of the order of the fine structure constant $\sim$ 1/137.

\section{Atomic target and photon energy spectrum for parity-even RENP}

\begin{figure}[htbp]
 \begin{center}
\includegraphics[width=8.0cm,bb=80 0 1300 1500, clip]{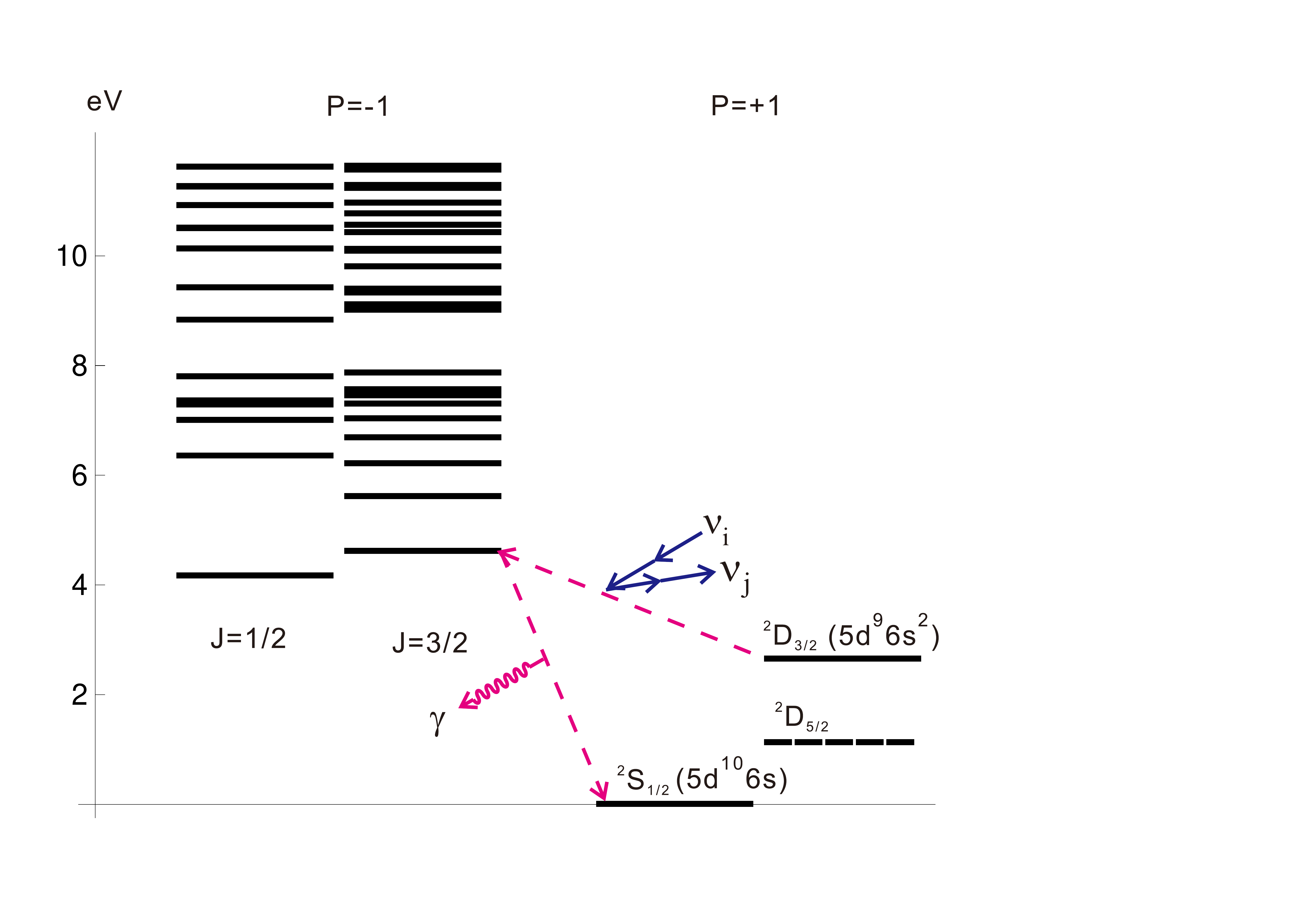}
   \caption{
Representative Au energy levels of different parities $P = \pm 1$ for RENP for 
${}^{2}D_{3/2} \to {}^{2}S_{1/2} + \gamma + \nu \bar{\nu}$.
The transition occurs via various $P = - 1$ intermediate states; among those, 15 $J = 1/2$ and 25 $J = 3/2$
states are taken into the actual calculation. The dashed lines show an example of considered transitions,
in which the transition amplitude is a product of a neutrino pair emission from $| e \rangle = {}^{2}D_{3/2}$ to $| p \rangle = {}^{2}P_{3/2}$ and an E1 radiative decay from $| p \rangle$ to $| g \rangle = {}^{2}S_{1/2}$.
}
 \label {p-even au level}
 \end{center} 
\end{figure}

We shall consider as a candidate atom of parity-even RENP, Au neutral atom or similar Cu.
Its level structure near the ground state is depicted in Fig. \ref{p-even au level}.
According to \cite{encapsulated atom} Au may well be isolated in encapsulated
fullerene from environments and maintain essential features of Au in vacuum.

In the RENP process of $| e \rangle \rightarrow |g \rangle + \gamma + \nu_i \bar{\nu}_j$,
neutrino-pair $ \nu_i \bar{\nu}_j$ of mass eigenstate is extremely difficult to detect,
hence we rely on the photon energy spectrum to extract properties of
massive neutrinos.
In actual RENP experiments we excite atoms in the ground state  by irradiating two
counter-propagating lasers of frequencies close to half the level spacing,
$\epsilon_{eg}/2$, which
have  very good energy resolutions much less than $O$(1 meV).
Along with a trigger laser we watch whether RENP occurs or not.
What is crucial for determination of energy resolution in experiments
is the laser frequency resolution rather than resolution of detected photon energy.
This way one can decompose amplitudes of neutrino pair emission into neutrino energy eigenstates instead of  flavor eigenstates $\nu_e$.
With finite neutrino masses the energy and the momentum conservation
of coherent RENP gives rise to six thresholds $ \omega_{ij} \,, i,j = 1,2,3$, which are shifted from
the massless neutrino value to  
$\omega \leq \omega_{ij} \,,  \omega_{ij} = \epsilon_{eg}/2 - (m_i+m_j)^2/(2\epsilon_{eg} )$
\cite{renp overview}.
Obviously, $\epsilon_{eg} > 2 m_3 $ must be obeyed to detect a neutrino-pair of largest mass $m_3 $.
Below,we will discuss in some detail how these threshold locations $\omega_{ij}$ are
manipulated by excitation lasers.

We consider Au excited state of $|e \rangle = 5d^9 6s^2$, decaying to the ground state
$|g \rangle = 5d^{10} 6s$.
Intermediate states of $|p \rangle = 5d^{10} np, 5d^{9} 6s np$ ($n=6,7,8,\cdots$) and 
$|p \rangle = np^5 5d^{10} 6s^2$ of a hole state ($n=3,4,5)$ appear in
the second order perturbation theory where $n$ is the principal quantum number.
Energy levels  of  these states are at
$ \epsilon_{eg} = \epsilon(5d^9 6s^2\, ^2D_{3/2} )  = 2.658  \,$eV and
$\epsilon(5d^{10} 6p\,^2P_{3/2}) = 5.105  \,$eV.

RENP amplitudes in the perturbation theory are
 for $\gamma_5$ parity-even RENP,
\begin{eqnarray}
&&
\frac{G_F}{\sqrt{2}} b_{ij} {\cal N}_{ij}{\cal M}(\omega)\equiv
\frac{G_F}{\sqrt{2}} b_{ij} {\cal N}_{ij} \times 
\nonumber \\
&& \hspace{2.5mm}
\sum_p \left(
\frac{\langle g | \vec{d}\cdot\vec{E}| p\rangle \langle p|\gamma_5 | e \rangle }{E_{g} - E_p +\omega }+ 
\frac{\langle g| \gamma_5| p\rangle \langle p| \vec{d}\cdot\vec{E}| e \rangle }{E_{e} - E_p - \omega } \right),
\label{atomic me 1}
\end{eqnarray}
and for $\vec{\alpha}$ parity-even RENP,
\begin{eqnarray}
&&
\frac{G_F}{\sqrt{2}} c_{ij} \vec{{\cal N}}_{ij} \cdot \vec{{\cal M}}(\omega)\equiv
\frac{G_F}{\sqrt{2}} c_{ij} \vec{{\cal N}}_{ij} \cdot 
\nonumber \\
&& \hspace{2.5mm}
\sum_p \left(
\frac{\langle g | \vec{d}\cdot\vec{E}| p\rangle \langle p|\vec{\alpha} | e \rangle }{E_{g} - E_p +\omega }
+ 
\frac{\langle g| \vec{\alpha}| p\rangle \langle p| \vec{d}\cdot\vec{E}| e \rangle }{E_{e} - E_p - \omega }
\right).
\label {atomic me 2}
\end{eqnarray}
Atomic parts of amplitudes, ${\cal M}(\omega)$ and $\vec{{\cal M}}(\omega) $, are
factored out of neutrino-pair emission parts, ${\cal N}_{ij}\,,\vec{{\cal N}}_{ij} $.
Neutrino-pair current $( {\cal N}_{ij}, \vec{{\cal N}}_{ij})  = \bar{\nu}_i \gamma_{\alpha}  (1 - \gamma_5)\nu_j   $ 
forms a 4-vector where $\nu_j , \bar{\nu}_i $ are neutrino plane wave functions of
definite helicities and momenta.
The two amplitudes given here are in principle comparable in magnitude, since
both $\gamma_5$ and $\vec{\alpha}$ matrices have small and large component
mixtures of 2-spinors.
Two types of parity-even RENP amplitudes never interfere when neutrino momenta
are integrated, and one can discuss two rates arising from
$\gamma_5$ and $ \vec{\alpha}$ separately.

The  idea  we use to  manipulate effectively threshold locations is
to exploit the phase memory at laser excitation of atoms.
With good quality of lasers of wave vectors, $\vec{k}_1, \vec{k}_2 $, excited atoms maintain 
imprinted phases of $e^{i (\vec{k}_1 + \vec{k}_2 ) \cdot \vec{x}}$.
This may be regarded as an initial atom having a momentum, $ \vec{k}_1 + \vec{k}_2  = \vec{p}_{eg}$,
hence with macro-coherence the
momentum conservation is changed to $\vec{p}_{eg}  -  \vec{k}  - \vec{p}_1 - \vec{p}_2= 0$
instead of $  \vec{k}  + \vec{p}_1 + \vec{p}_2= 0$
where $\vec{p}_j\,, j = 1,2$ momenta of neutrino pair and $ \vec{k}$ the momentum of observed photon.
This further changes the effective mass of the neutrino pair to $\sqrt{s}$ 
with $s=q^2$ and $q \equiv (q_0,\vec{q})=(\epsilon_{eg} - \omega , \vec{p}_{eg} - \vec{k})$,
which may be  smaller than  the value without the imprinted phase, if
$|\vec{p}_{eg} -   \vec{k} | > | \vec{k} | $.
We term this case boosted RENP \cite{boosted renp}.

Spectrum rate formula of boosted RENP for
$\gamma_5$ parity-even case is given by 
$\Gamma=\sum_{ij} \Gamma_{ij}$, with 
\begin{eqnarray}
&&
\Gamma_{ij}= \frac{2 G_F^2}{\pi} |{\cal M}(\omega)|^2  n^2 V \times
\nonumber \\ && \hspace{0mm}
\int_{E_-}^{E_+} dE_1 J_{ij}(E_1, \epsilon_{eg} - \omega - E_1)  \theta ( \sqrt{s} - m_i - m_j )
\,, 
\label {rate formula}
\\ && \hspace{0mm}
J_{ij}(E_1, E_2) = \frac{1}{|\vec{q}|} \Big[ |b_{ij}|^2 
\Big( 2E_1 E_2 -\frac{s-m_i^2-m_j^2}{2} \Big)
\nonumber \\ && \hspace{20mm}
+ \delta_M\;(\Re b_{ij}^2)\;  m_i m_j \Big],
\label{rate kernel}
\end{eqnarray}
where 
$E_{\pm}$ is the maximum and the minimum neutrino energy, 
and $\delta_M = 1(\delta_M = 0)$ is applied for the Majorana (Dirac) neutrino.
The trigger laser field $\vec{E}$ in ${\cal M}(\omega)$ in Eq.(\ref{rate formula}) should be understood 
as $\rho_{eg} \vec{E} $ where $\rho_{eg}$ is coherence developed in target atoms. 
It is convenient to introduce a dimension-less parameter $\eta$ defined as 
$|\langle \rho_{eg}\vec{E} \rangle|^2= \omega n \eta/2$\, \cite{boosted renp}.
The factor $ \eta$ is coherence averaged over target atoms,
typically of order
$10^{-3} \sim 10^{-6}$ from computations of \cite{psr}.
In numerical computations below
we use parallel or anti-parallel configuration in which
the initial wave vector $\vec{p}_{eg} $ is taken along $ \vec{k}$,
with its magnitude by $p_{eg}  = r \epsilon_{eg} \,, - 1 \leq r \leq 1 $.
For $ \vec{p}_{eg}$ parallel to $  \vec{k}$ the neutrino-pair production thresholds are at
$\omega_{ij} = \frac{ 1+r}{2} \epsilon_{eg} - \frac{ (m_i+m_j)^2}{ 2 (1-r)\epsilon_{eg} } $.
All six thresholds can be observed if the condition
$r^2 < 1- 4 (m_1/\epsilon_{eg})^2 $ with $m_1 $ is the smallest neutrino mass
is satisfied.

\begin{figure}[htbp]
 \begin{center}
\includegraphics[width=8.0cm,bb=150 100 1500 1000, clip]{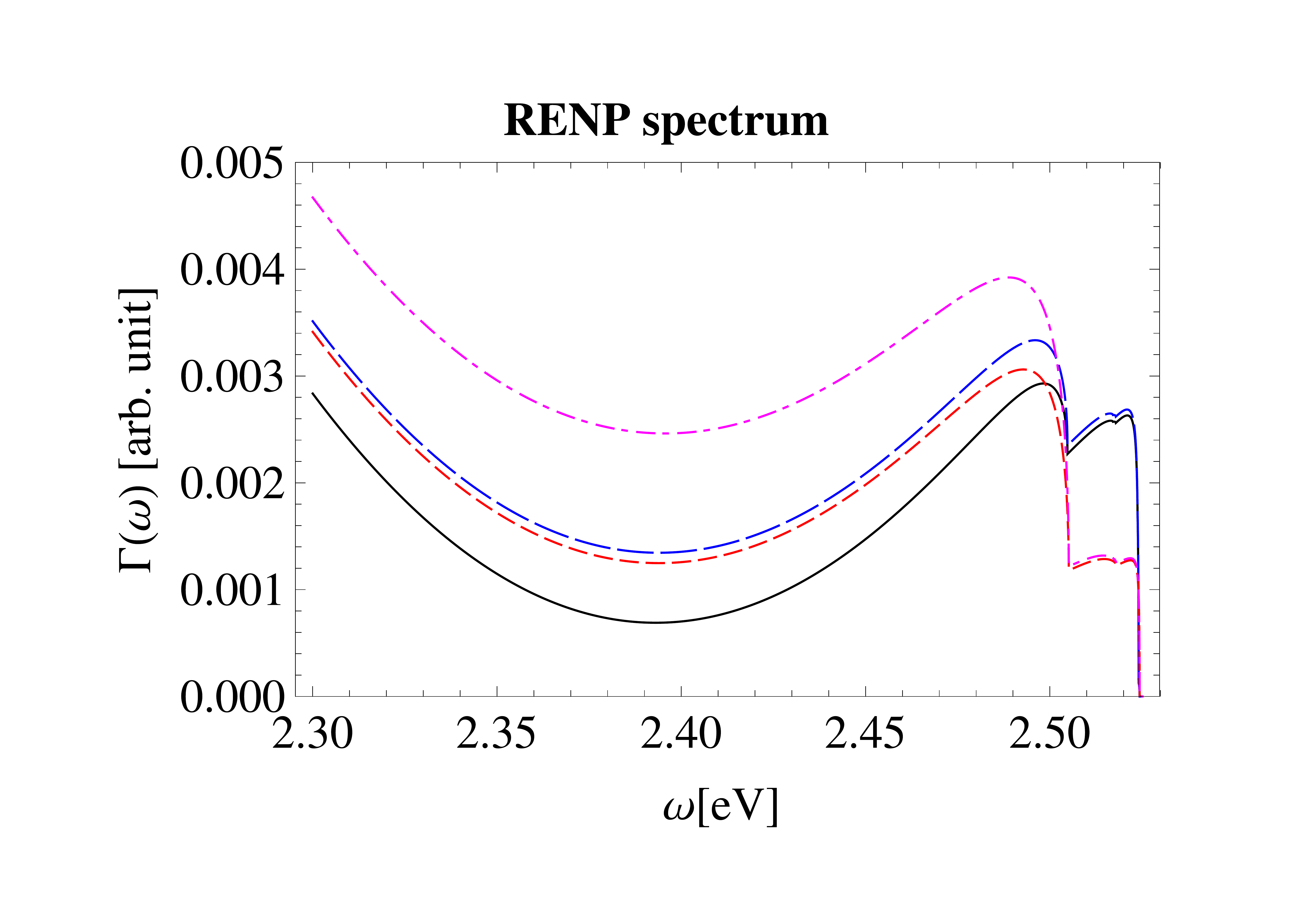}
   \caption{ 
Comparison of Dirac and Majorana spectra for the inverted (IO) and normal ordering (NO) cases with a parallel initial phase memory of $r=0.9$:
NO Dirac rate in solid black, NO Majorana in long-dashed blue, IO Dirac in short-dashed red,
and IO Majorana in dash-dotted magenta for
the smallest mass of 10 meV and assuming vanishing CP violating phases.
Absolute rate value is obtained by multiplying $10^{-6} $sec$^{-1}$ for $n= 10^{22} {\rm cm}^{-3}, V= 10^2 {\rm cm}^3,
\eta = 1 $.
}
   \label {renp spectrum shape ino 2}
 \end{center} 
\end{figure}

The bench mark rate unit of parity-even RENP is
estimated as a product of squared atomic matrix elements of order $10^{-3} $ in the atomic unit,
the neutrino-pair integral of order $O(1 \sim 10^{-2})$ eV$^{2}$ from Fig. \ref{renp spectrum shape ino 2},
and $G_F^2/2\pi$, which gives 
$\sim10^{-6}$sec$^{-1}$,
taking $ n= 10^{22} {\rm cm}^{-3}\,$,
$  V=10^2 {\rm cm}^3\,$ and $\eta = 1$.


\section{Details of atomic calculation}

We now turn to atomic parts of amplitude ${\cal M}(\omega)\,, \vec{{\cal M}}(\omega) $. 
Eqs (\ref{atomic me 1}) and  (\ref{atomic me 2}) were evaluated for neutral gold, where 
we selected the $5d^{9}6s^{2}$($^{2}D_{3/2}$) and $5d^{10}6s$($^{2}S_{1/2}$) 
as the $|e\rangle$ and $|g\rangle$ states, 
respectively. For the $|p\rangle$ states in Eqs.(\ref{atomic me 1}) 
and  (\ref{atomic me 2}), fifteen odd-parity J=1/2 
valence excited states and twenty-four odd-parity J=3/2 states were considered. 
In addition, $np^{5} (n=3,4,5)$ odd-parity J=1/2 and 3/2 core-excited states were 
also included in the calculation. 
The $2np^{5}$ core-excited states were not included 
because of the program restriction, but its effect on the results 
are expected to be small since the excitation energies are much higher 
than the other valence or core-excited states.

The wave functions for these atomic states were calculated based on the multi-configuration 
Dirac-Hartree-Fock method and the relativistic configuration interaction method \cite{grant} implemented 
in the GRASP2K package \cite{grasp1}, \cite{grasp2}.
The wave functions, i.e., the atomic state 
functions, for these atomic states were represented by linear combinations of configuration 
state functions, which were constructed from single-particle Dirac orbitals. 
The single-particle Dirac orbitals were determined by the multi-configuration 
Dirac-Hartree-Fock method, whereas the expansion coefficients of the linear combination 
were calculated by the relativistic configuration interaction method. 
In the first step of the calculation, using the multi-configuration Dirac-Hartree-Fock method, 
the $|g\rangle$
 state was represented by the linear combination of the $5d^{10}6s$ and $5d^{9}6p^{2}$ configurations,  
 and $|e\rangle$
by the $5d^{9}6s^{2}$ and $5d^{9}6p^{2}$ configurations, 
while $|p\rangle$ by the $5d^{9}6s6p$, $5d^{10}6p$ and $5d^{9}6p6d$ configurations. 
The inner orbitals were treated as inactive occupied core-orbitals. These configurations will be referred as 
multi-reference set hereafter. Then the self-consistent field procedure was applied to optimize both the expansion 
coefficients for the multi-reference configurations and the Dirac orbitals simultaneously. 
To correct for dynamic electron correlation effects, separate calculations for the $|g\rangle$, $|e\rangle$, and $|p\rangle$ states 
were performed with expansions including CSFs obtained by single (S) and double (D) excitations from 
the multi-reference set to active sets of orbitals up to $8s$, $10p$, $8d$, $6f$, $6g$ and $6h$. 
The core-valence and core-core correlations involving the $5d^{10}$ core were included in addition to 
the valence correlations. In the last step, relativistic 
configuration interaction calculations were performed in which the Breit interaction \cite{grant}, 
vacuum polarization \cite{vac pol}, and self-energy \cite{vac pol} were considered. 
The transition matrix elements of the electric dipole, $\gamma_5$, and $\vec{\alpha}$ operators 
in Eqs.(\ref{atomic me 1}) and  (\ref{atomic me 2})
were calculated by the reduced matrix elements, evaluated by the bi-orthogonalized atomic state functions, 
using the Wigner-Eckart theorem \cite{grasp2}.  

Results show that the zero-th component of axial vector
contribution in the neutrino-pair emission vertex is much larger (by $\sim O(10)$ at $\omega \leq 3 $eV)
 than the spatial component of vector contribution.
We shall use the fitting function \cite{atomic amp fitting} arising $\gamma_5$ vertex alone
for simplicity in the following.

\color{black}

\begin{figure}[htbp]
 \begin{center}
\includegraphics[width=8.0cm,bb=100 0 1600 1000, clip]{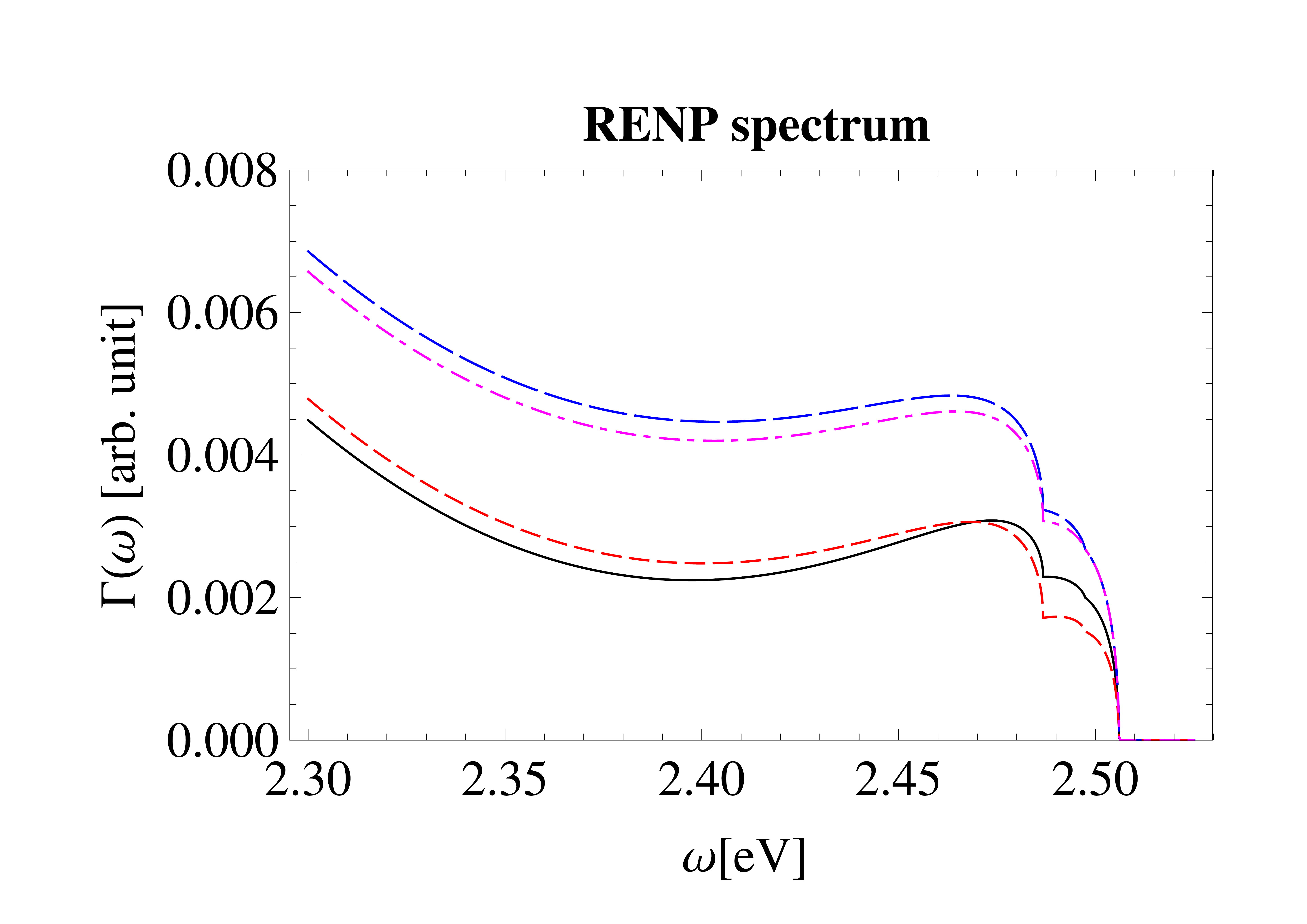}
   \caption{ 
CPV phase $(\alpha,\beta-\delta)$ dependence of spectrum:
Comparison of normal ordering (NO) spectra with a parallel initial phase memory of $r=0.9$:
Dirac in solid black, Majorana $(\alpha  =0, \beta-\delta=0)$
 in long-dashed blue, Majorana $(\alpha  =\pi/2, \beta-\delta=0) $ in short-dashed red,
and Majorana $(\alpha=0,\beta-\delta =\pi/2) $ in dash-dotted magenta for
the smallest mass of 50 meV.
}
   \label {renp spectrum shape cpv}
 \end{center} 
\end{figure}

\section{Results of RENP photon spectrum}

We present numerical results in 
Figs. \ref{renp spectrum shape ino 2} and \ref{renp spectrum shape cpv} in the presence of an
initial phase memory $r$.
For these calculations of spectrum shapes we take
squared neutrino mixing factors $ |U_{ei}|^2 \,$( i = 1,2,3), 
hence $|b_{ij}|^2 $, calculated from neutrino oscillation data \cite{pdg}.
CP violating (CPV) phases, $\delta$ for Dirac neutrino and
$\delta, \alpha, \beta$ for Majorana neutrino, appear in
$ U_{ei} U^*_{ej}\, ( i\neq j) $, and they are experimentally unknown.
Other parameters determined by neutrino oscillation experiments
are two mass squared differences, $\delta m^2_{ij} $.
We assume in our analysis a particular smallest neutrino mass,
and calculate other masses from these differences.
Spectrum results are shown  both for normal ordering (NO) neutrino mass pattern
and inverted ordering (IO) \cite{pdg}, which is experimentally undetermined so far.

A few comments we would like to make are

(1) the smallest neutrino mass determination is easiest at $ \omega_{12} $ threshold in which
the weight factor $|b_{12}|^2 = 0.405$ is largest, 

(2) the Majorana/Dirac distinction is easiest after $ \omega_{33} $ threshold opens with
$|b_{33}|^2 = 0.227$,

(3) unlike parity-odd RENP, parity-even RENP in the present work gives larger Majorana pair emission rates 
than Dirac pair emission, 

(4) all other thresholds have much smaller (by $O(1/10)$) weights, but
they are in principle detectable with finer resolutions of high statistics data,
since two thresholds, $ \omega_{12}, \omega_{33} $, making up most of the summed weights, $\sum_{ij} | b_{ij}|^2 = 3/4$,

(5)
The term $\Re b_{ij}^2$ in Eq.(\ref{rate kernel}) gives CPV phase dependences on $\alpha$ and $\beta-\delta$. 
It is largest for $\alpha$,
since $ b_{12} \propto e^{- i \alpha}$ and the phase independence of $b_{33} $.
Dependence on $\beta-\delta$ is less sensitive due to
their appearance in $U_{\rho i} U_{\rho j}^*\, ( i \neq j)$ of flavors $ \rho = \mu, \tau$.

The most serious background arises for macro-coherence involving QED processes,
and for parity-even RENP discussed in the present work
the background is macro-coherent four-photon process called McQ4 \cite{mcqn}.
How to suppress this background is discussed in \cite{photonic crystal}.

Unambiguous detection of atomic neutrino far above six thresholds is expected to give
an overall neutrino mass scale, although the smallest neutrino mass
measurement needs more dedicated efforts.
In this sense the discovery of neutrino-pair production in atomic de-excitation
is surely a breakthrough in neutrino physics.

\vspace{0.5cm}
In summary,
we proposed a new experimental method of measuring the smallest neutrino mass
and determining whether neutrinos are of Majorana or Dirac type,
and presented results of how to extract neutrino properties by measuring emitted
photon energy spectrum.
Isolated neutral atoms in solid environment is necessary to prepare large enough targets in actual experiments.

\section*{Acknowledgement}
This research was partially
 supported by Grant-in-Aid 16K05307(MT) and 16H00939(MT), 15H02093 (NS), 17H02895 (MY),  from the
 Ministry of Education, Culture, Sports, Science, and Technology of Japan. 
JE and PJ have been supported by the Swedish Research Council (VR)
under contract 2015-04842. 
JE's visit to Tokyo Institute of Technology to initiate collaboration on the work reported 
here was supported by a grant from MEXT Japan under the auspices of The Program for Promoting 
the Enhancement of Research Universities.


\begin{thebibliography}{99}
\bibitem{pdg}
Particle Data Group Collaboration, M. Tanabashi
et al., Phys. Rev. {\bf D 98}, 030001 (2018).



\bibitem{renp overview}
A. Fukumi et al.,
Prog.\ Theor.\ Exp.\ Phys.\ (2012) 04D002,
and earlier references cited therein.




\bibitem{psr exp}
Y. Miyamoto et al., PTEP, 113C01 (2014).
Y. Miyamoto et al., PTEP, 081C01 (2015).
T. Hiraki et al., arXiv:1806.04005 [physics.atom-ph],
and J. Phys. {\bf B 52}, 045401 (2019).



\bibitem{encapsulated atom}
S. Saito, Y. Sugaya, and M. Toyoda.
Private communication.

\bibitem{boosted renp}
M. Tanaka, K. Tsumura, N. Sasao, and M. Yoshimura, Phys. Rev. {\bf D96} 113005 (2017).

\bibitem{psr}
M. Yoshimura,N. Sasao, and M. Tanaka, Phys. Rev. A 86, 013812 (2012).
The quantity $\eta$ was introduced for the first time in reference [2].

\bibitem{grant}
I.P. Grant,
{\it Relativistic Quantum Theory of Atoms and Molecules},
Springer, New York (2007).

 
\bibitem{grasp1}
P. J\"onsson, X. He, C. Froese Fischer and I.P. Grant, Comput. Phys. Commun.{\bf 177}, 597 (2007).

\bibitem{grasp2}
 P. J\"onsson, G. Gaigalas, J. Bieron, C. Froese Fischer and I.P. 
Grant,  Comput. Phys. Commun. {\bf 184}, 2197 (2013).

\bibitem{vac pol}
P.J. Mohr, G. Plunien, and G. Soff,
Phys. Rept. {\bf 293}, 227 (1998).


\bibitem{atomic amp fitting}
{
The fitting formula of amplitude $|{\cal M}(\omega) |^2 $ used in spectrum computations is
$
a_0 + a_1(x-0.05) + a_2(x-0.05)^2 + a_3(x-0.05)^3 + a_4(x-0.05)^4
\,
$, with 
$a_0 = 0.00103776,\,$
$a_1 = -0.00700889,\,$
$a_2 = 0.136222,\,$
$a_3 = -1.08185,\,$
$a_4 = 0.846076\,$ 
in terms of the atomic unit of $x \simeq \omega/ 27.2 {\rm eV} $.
}

\bibitem{mcqn}
M. Yoshimura, N. Sasao, and M. Tanaka,
 PTEP, 053B06 (2015).

\bibitem{photonic crystal}
M. Tanaka, K. Tsumura, N. Sasao, and M. Yoshimura,
 PTEP, 043B03 (2017).

\end{thebibliography}
\end{document}